# Carbon financial system construction under the background of dual-carbon targets: current situation, problems and suggestions

Yedong Zhang　　Han Hua＊

Table of content

---

＊Yedong Zhang，Associate Professor at Law School of Shenzhen University and Research Associate at the Shenzhen University Institute of Science, Technology, and Law.
Han Hua, Shanghai University of Political Science and Law, hhuaview@163.com (H. Han).

**Abstract**

Under the guidance of the dual-carbon target, the development of the carbon financial system is of great significance to compensate for the gap between green and low-carbon investment. Considering the current state of the development of carbon financial system, China has initially formed a carbon financial system, including participants, carbon financial products and macro and micro operation structures, but the system is still in the initial development stage. Given the current restrictions on the development of carbon finance, it can be seen that there are still problems such as unreasonable economic structure, insufficient market construction, single product category, low utilization rate and urgent construction of relevant judicial guarantee system. Therefore, the system should be improved at the economic level and the legal level. The economic level includes adjusting the layout of economic development structure, strengthening the construction of market infrastructure, encouraging the diversification of carbon financial products and strengthening publicity and education promotion strategies. The legal level includes improving the top-level design, formulating judicial interpretation to promote carbon financial trading, promoting commercial law amendment, and promoting the linkage mechanism between specialized environmental justice and carbon finance and other safeguard measures. Finally, improving the carbon finance system is required to promote and protect the orderly development of carbon finance. To promote the reform of the pattern of economic development, the concept of ecological and environmental protection in the financial sector needs to be implemented to form an overall pattern of pollution reduction, carbon reduction and synergistic efficiency improvement.

ecological environmental protection

# 1 Foreword

At present, climate change is an important issue that mankind needs to face, and climate investment and financing are on the rise. The carbon finance system plays a crucial role in supporting the transition towards a low-carbon economy by channeling financial resources into green projects.[1] It supports initiatives such as positive energy districts, which focus on producing more renewable energy than consumed. This system encourages the development of energy-efficient infrastructures and low-carbon technologies, aligning with global sustainability goals. Through financial instruments like carbon credits and green bonds, the system helps finance projects that transition urban areas towards sustainable, carbon-neutral energy solutions, fostering the broader achievement of carbon neutrality.[2] It involves tools such as green credit, green bonds, and carbon markets to promote sustainable development. By incentivizing investments in clean energy and low-carbon technologies, carbon finance helps optimize energy efficiency in sectors like housing, industry, and transportation. Its integration with environmental policies facilitates the achievement of carbon neutrality goals and supports long-term sustainable economic growth.[3] In the next 30 years, the scale of green and low-carbon investment that China needs for achieving carbon neutrality should be hundreds of millions of yuan. The government's financial funding level is far from meeting such a huge funding gap. Therefore, promoting the development of the carbon financial market and supporting the legal system are urgent practical problems that need to be solved at present. The national carbon emission trading market was officially launched at the end of June 2021. China currently has eight local pilot carbon markets, and the national carbon trading platform was finally settled in Shanghai. According to relevant professionals, the move means that the carbon market and carbon tax will be implemented by different industries in the future.[4] This analysis clearly indicates that the carbon market construction measures will be further strengthened and improved in the future. It is also a huge input into the development of carbon finance. Because the development of carbon finance depends on the construction and improvement of the carbon market infrastructure, the move sends a key signal to the market to achieve the two-carbon target in the future. Further financialization of the carbon market will become an important part and wind vane for the construction of China's financial system to lead further improvement and development of the green financial system. Realizing the transformation of production mode and solve the environmental problems at their root causes, and providing financial support for the harmonious development of man and nature and the construction of socialist ecological civilization will provide China with unique Chinese wisdom and solutions for building a community with a shared future between man, nature and ecology.

---

[1] Carbon finance (Carbon Finance), also known as low-carbon finance, generally refers to the financial system, related investment and financing activities that serve to reduce greenhouse gas emissions. The promulgation of the United Nations Convention on Climate Change (United Nations Framework Convention on Climate Change, UNFCCC) and the Kyoto Protocol (Kyoto Protocol) have laid the foundation for the construction of the carbon financial system. Carbon finance is the essence for coping with global warming, achieving green low carbon economic development, increasing international cooperation space, promoting the construction and perfect, promoting the development of high energy consumption enterprise transformation, promoting the orderly development of low carbon industry and eventually realizing the comprehensive optimization of national energy structure; moreover, it is ultimately conducive for improving the national and regional carbon market system, promoting the territory of green finance extension, reflecting the image of the head of our country, and enhancing the international competitiveness and influence.
[2] Akhatova, Ardak, et al. "Techno-economic aspects and pathways towards positive energy districts: status quo and framework conditions as a basis for developing techno-economic pathways in selected case studies." (2020).
[3] Congxiang, Tian, et al. "Establishing Energy-efficient Retrofitting Strategies in Rural Housing in China: A Systematic Review." Results in Engineering (2024).

    However, we must be clearly aware that under the background of the current macrofinancial state and financial reform, China's carbon financial development still faces many challenges in investment and financing, and there is a large gap in terms of the national green and low-carbon development goals and the requirements for the dual-carbon goals. The carbon finance system is essential for guiding investments into environmentally sustainable projects. It includes mechanisms like green credit, carbon

---

[4] They are Shenzhen, Shanghai, Beijing, Guangdong, Tianjin, Hubei, Chongqing and Fujian.

trading, and green bonds, aimed at reducing carbon emissions and promoting green development. This system encourages the allocation of funds to low-carbon sectors, helping to optimize industrial structures and drive technological innovation. By leveraging financial tools to support the green economy, carbon finance aligns with national carbon neutrality targets, fostering high-quality economic growth and contributing to sustainable development in regions like the Shaanxi Pilot Free Trade Zone.[5] In a green financial system that has not yet been established and it is not perfect, the green financial development model is not clear (top to bottom or bottom-up), there are green small and medium-sized enterprise financing difficulties, climate investment and financing system framework are not perfect, and green financial standards and carbon neutral standard are not coordinated; the challenge of the new development pattern with the epidemic situation background, green finance and pollution prevention of effective synergies and cohesion have not been effectively solved ,therefore, from the micro perspective of the construction of carbon financial system, it is perhaps possible to find effective countermeasures to resolve the above problems. Hence, this article analyzes and demonstrates the current state, problems and corresponding countermeasures of the carbon financial system under the background of a two-carbon target.[6]

## 2 Status quo: Carbon financial system under the background of the two-carbon target

As a branch of the green financial system, the mission of carbon finance is to give full play to the resource allocation, risk management and market pricing functions of finance to support low-carbon development, internalize the externalities of carbon emissions through factor allocation and trading mechanisms, and eliminate the embarrassing situation of the administration and public welfare. The legal principle behind carbon finance lies in the profound change in human life and production and the gradual reform of the financial industry in terms of the real economy.[7] The carbon finance system includes tools like carbon credit ratings, which are critical for promoting environmentally sustainable business practices. By evaluating enterprises' carbon performance through energy data, carbon credits provide a mechanism for companies to offset their emissions. These ratings incentivize businesses to adopt green technologies and improve energy efficiency. Additionally, they enable the creation of a financial market for carbon trading, supporting the broader transition to a low-carbon economy. This system not only helps reduce carbon footprints but also enhances corporate sustainability, aligning with global carbon neutrality goals.[8] In the next 40 years, China will need an investment to achieve the carbon neutral target, and the financial sector has broad development prospects in supporting the dual-carbon target.[9] As the core participants in the financial field, financial institutions are not only the best lever to leverage social funds but also an important participant in climate investment and financing. To ensure the orderly participation of financial institutions in the carbon financial market, the People's Bank of China and the Ministry of Ecology and Environment jointly issued the Guidelines on Promoting Investment and Financing on Climate Change in October 2020, and these guidelines require market orientation. We will give full play to the decisive role of the market in climate investment and financing, better leverage the role of the guidance of the government, and effectively leverage the role of the financial institutions and enterprises as innovation players in the models, mechanisms, and financial instruments.[10] In particular, the document proposed two major measures to improve the carbon emission trading mechanism and strengthen financial policy support in terms of focusing on the role of financial institutions and guiding more social funds to the climate field.[11] It can be seen that the two important aspects of the future development of carbon finance are the economic system and the legal system.[12] From the research logical starting point, to find the current problems of the carbon financial system, it is necessary to make a specific analysis of the current state of the system operation. Therefore, the actual operation of the carbon financial and economic system and the legal system in China will be discussed in detail next.

---

[5] Zhang, Peng, and Zhongli Li. "Studies on How to Innovate the Green Finance System to Support the China (Shaanxi) Pilot Free Trade Zone's High-Quality Development."
[6] See An G (2021) Discussion on green finance innovation path under the carbon neutralization goal. South Finance, Baidu (2021) https://baijiahao.baidu.com/s?id=1697897198974347585&wfr=spider&for=pc.

---

[7] Wu, X. "To Strengthen and Expand the Carbon Financial Market." China Business Daily, 2021.
[8] Meng, Shiyu, and Junping Yin. "Evaluation and Application of Enterprise Carbon Credit Rating Based on Energy Data." Proceedings of the 4th International Conference on Informatization Economic Development and Management, IEDM 2024, February 23–25, 2024, Kuala Lumpur, Malaysia. 2024.
9  Liu, L. "Carry Out Carbon Finance Work, Help Carbon to Reach the Peak and Carbon Neutral." People's Daily, 2021b.

## 3 The current situation of the carbon financial and economic system

The carbon financial market is the financialized carbon market. It is the mainstream of the development of the European and American carbon markets and the future development direction of China's carbon market. The carbon finance system, integrated with digital finance, plays a significant role in reducing urban carbon emissions. By leveraging digital technologies, it enhances the efficiency of financial transactions, enabling more accessible investments in low-carbon projects. Digital finance facilitates the development of green bonds, carbon credits, and eco-friendly investment products, driving capital towards sustainable ventures. This system not only supports carbon reduction goals but also promotes the transformation of urban industries towards sustainability, thereby contributing to the achievement of carbon neutrality targets under the dual carbon goals.[13] Relying on carbon quotas and project emissions reduction, there are two kinds of carbon asset development financial tools, mainly trading tools (carbon futures, carbon forward, carbon swap, carbon options, etc. ), financing tools (carbon pledge, carbon repurchase, carbon custody, etc.) and support tools (carbon index and carbon insurance, etc. ); these three kinds of tools can help market participants more effectively manage carbon assets, provide diversified trading methods, improve market liquidity, hedge future price volatility risk, and provide carbon hedging. The carbon market hierarchy includes the macro level of the carbon trading regulation system (ETS) and the micro level of the secondary trading market, financing service market and support service market. The secondary market is the core, and it is divided into floor trading and otc, and the macro framework and the transition part are the primary market. The maturity of carbon financial market development can be measured from the perspective of market efficiency. One part is the efficiency of resource allocation, and the other part is the efficiency of market operation. Resource allocation includes the effectiveness of the and the stability of the carbon price. The efficiency of market operation includes market liquidity and the influence of the carbon price. The development of the carbon financial market is of great significance to improving the carbon market mechanism, enriching the green financial system and striving for international carbon pricing power.

---

10 Mee (2020) http://www.mee.gov.cn/xxgk2018/xxgk/xxgk03/202010/t20201026_804792.html.
11 Ge X (2021) The practical dilemma and outlets for financial institutions to participate in climate investment and financing business. Southwest Finance.
[12] Its essence is that the economic base determines the superstructure. The development of the real economy cannot be achieved without financial support. Finance, like the human blood, supports the sustainable development of the real economy. Therefore, the development of carbon finance is just like the source of the realization of double carbon goals. Moreover, carbon finance must be steady and orderly, hence, an absence of strict regulatory framework will lead to falling into a disorderly development. As with other types of financial markets, carbon finance must prevent economic virtualization, hollowing out, and gamification. We must always stay true to our original aspiration to achieve the two-carbon goal, hence, the financial regulatory policy alone is not enough. It is imperative to achieve the top-level design and refinement of the carbon financial legal system.

---

*3.1 Participating subjects*

       The participants of the carbon financial market include four parties(see table 1): trading parties[14], third-party intermediaries[15], fourth-party platforms[16] and regulatory authorities.[17]The specific classification, role, influence and main motivations of the participants in the carbon financial market are shown below.

---

[14] Refers to the buyers and sellers who directly participate in the carbon financial market trading activities, mainly including emission control enterprises, emission reduction project owners, carbon asset management companies, carbon funds, financial investment institutions and other market entities. In the spot trading stage, market players are often dominated by control enterprises, supplemented by carbon asset management companies and financial investment institutions; in the derivatives trading stage, financial investment institutions, especially market makers and brokers, will become the main providers of market liquidity.
[15] Refers to the professional institutions that provide various auxiliary services for market entities, including monitoring and verification institutions, certification institutions, consulting companies, evaluation companies, accountants and law firms, as well as the institutions that provide financing services to both parties.
[16] Refers to the service institutions that provide public infrastructure for the market parties to carry out transaction-related activities, mainly including registers and exchanges. Among them, exchanges not only provide trading places, trading rules, trading system, trading matching, clearing delivery and information services but also undertake the daily supervision functions of some front-line trading activities in the market.
[17] It refers to all kinds of competent departments that manage and supervise the compliance and stable operation of the carbon financial market, mainly including industry competent departments, financial regulatory departments and fiscal and taxation departments.

**Table 1.** Specific classification, role, influence and main motivations of the participants in carbon financial markets

| Classification | Role | Influence | Motivations |
|---|---|---|---|
| Both parties to the transaction | Control and discharge enterprises | Market dealing. Improve energy efficiency and reduce energy consumption, and drive the whole society to achieve the emission reduction targets through individuals in the real economy. Minimum cost emission reduction through intersubject transactions. | Complete emission reduction targets (performance). Sell low and sell high to achieve a profit. |
| | Emission reduction project owner | Provide the required emission reduction to reduce the cost of performance. Promote emission reduction efforts for subjects not included in the trading system and other industries. | Emission reduction generated by selling emission reduction projects to achieve economic and social benefits. |
| | Carbon Asset Management, Inc. | Offering consultation services. Invest in carbon financial products to enhance market liquidity. | Buy low and sell high, to achieve a profit. |
| | Carbon fund and other financial investment institutions | Enrich the trading products. Attract investment money in Terms of Enhancing market liquidity. | Expand your business and profit. |
| Third party intermediary | Monitoring and certification agencies | The "three" principles of guaranteeing carbon credit (measurable, reportable and verifiable). Maintain the effectiveness of market transactions. | Admidia expansion business. |
| | Others (e. g. consulting firms, appraisal companies, accountants and law firms) | Offering consultation services. Irene carbon asset assessment. Carbon-trading-related audits. | Admidia expansion business. |
| Fourth party platform | Registration agency | Register for carbon quotas and other carbon credit indicators permitted by regulations. Standardize market trading activities and facilitate | Ensure the standardization and security of market transactions. |
| | | regulation. | |
| | Trading platform | The collection and release of transaction information. Reduce transaction risk and reduce transaction costs. Vacation price discovery. Enhance market liquidity. | Attract buyers and sellers to trade, enhance market liquidity and benefit from it. |
| Supervision department | Carbon trading regulator | We will formulate regulations on the carbon emission reduction quota trading market, and exercise regulatory power in accordance with the law and regulations. To supervise the listed trading varieties and supervise the specific implementation of the trading system and trading rules. Supervise the trading activities of the market. To supervise and inspect the information disclosure of market transactions. Cooperate with relevant departments to investigate and punish violations of laws and regulations to maintain the health and stability of the market. | Standardize the market operation through market supervision. Promote energy conservation and emission reduction through market mechanism operation. |

*3.2 Carbon financial products*

As mentioned above, the carbon financial market is the financialized carbon market. From the perspective of the product spectrum, carbon financial products are mainly the mapping of mainstream financial products in the carbon market, and these can be divided into three categories: trading tools, financing tools and support tools. Trading instruments include carbon futures, carbon options, carbon forwards, carbon swaps, carbon index trading products, carbon asset securitization, carbon funds, carbon bonds (especially carbon neutral bonds) and other products.[18] Financing instruments include carbon pledge, carbon repurchase and carbon custody products.[19] Carbon market support tools include the carbon index, carbon insurance and other products.[20]

*3.3 Hierarchy*

As an artificially created market, especially one that is created and driven by policies, the hierarchy of the carbon market can be discussed from two aspects: macro framework and microstructure.

---

18 Carbon futures is a contract with carbon emission right quota and project emission reduction as the spot contract. The basic elements include trading platform, contract scale, margin system, quotation unit, minimum trading scale, minimum/maximum volatility, contract maturity date, settlement method, clearing method and so on. Carbon option, essentially a trading right, refers to the right to sell or purchase a certain amount of the subject matter after paying a certain amount of right to the seller who shall pay the right, and the subject matter is spot or futures for carbon emission right. Carbon forward trading refers to the trading method in which buyers and sellers agree to buy and sell a certain amount of carbon assets such as quotas or project emission reduction at a certain price at a certain period in the future. Carbon swap is based on the carbon emission right as the subject matter. Both parties determine the transaction at a fixed price, and agree to complete the reverse transaction corresponding to the fixed price transaction at the current market price at some time in the future. Finally, only the price difference between the two transactions should be settled in cash. Carbon index trading products refers to the development of corresponding carbon index trading products with carbon market index as the subject matter. In addition, the future usufruct of carbon quota and emission reduction projects can be used as supporting assets for financing through securitization, among which securities securitization is carbon fund and bond securitization is carbon bond.
19 Carbon pledge refers to the debt financing guaranteed by carbon assets such as carbon quota or project emission reduction. The borrower edges the carbon assets to creditors such as banks or securities firms to obtain certain discount financing, and then pays the principal and interest at maturity. Carbon repurchase refers to a short-term financing arrangement in which carbon quota holders sell quotas to other institutions and agree to repurchase the quotas sold at an agreed price for a certain period of time. During the term of the agreement, the transferee may dispose of the carbon quota by itself. Carbon custody (carbon borrowing) refers to the activities in which a party entrusts its carbon assets to a professional carbon asset management institution to maintain and increase its value; for a carbon asset management institution, carbon custody is actually a carbon melting tool.
20 Carbon index is an indicator that reflects the overall price of the carbon market or the price changes and trends of a certain type of carbon assets. It is a yardstick to describe the scale and change trend of carbon trading. Carbon index is not only an important observation tool of the carbon market but also the basis for the development of carbon index trading products. Carbon insurance is a guarantee tool to avoid risks in the development process of emission reduction projects and ensure that the project emission reduction is delivered in full and on schedule. It can reduce the investment risk or default risk of both parties of the project and ensure the smooth progress of the project investment and transaction behavior.

The carbon finance system plays a vital role in promoting the green economy by directing financial resources towards sustainable agricultural practices. It includes mechanisms such as green credit and carbon finance, which support eco-friendly initiatives in the agricultural supply chain. These financial tools help reduce environmental impacts by encouraging sustainable farming practices and resource-efficient technologies. By evaluating the development level of agricultural supply chain finance, the system aims to optimize investment, promote low-carbon solutions, and align agricultural activities with environmental sustainability goals.[21] The macro level mainly refers to the carbon trading system (ETS) under the regulation of government policies; the micro level includes the secondary trading market, financing service market and support service market as the core, and these are divided into floor trading and over-the-counter trading; and the transition link between the macro framework and microstructure is the primary market.

*3.3.1 Macrolevel system framework*

The macro system framework includes the key elements of the ETS, the carbon pricing area and its connection(see table 2). First, the key elements of an ETS include coverage, total setting, quota allocation, offset mechanisms and performance supervision (International Carbon Action Partners (ICAP) 2016). The second is the carbon pricing area and its connection. According to the geographical jurisdiction covered by the ETS, the carbon market can be divided into a regional market, national market and international market, forming carbon pricing areas at different levels.[22] How to connect different carbon pricing zones is of great significance to the global synergy and strengthening the action against climate change. The connection has two dimensions: one is the scope, including the international and domestic levels and different carbon pricing areas, and the other is the object, including the quota level and the project level. There are already rich and colorful specific practices.

---

21 Lu, Jiaxuan, and Hangyu Cao. "Research on the Evaluation of Agricultural Supply Chain Finance Development Level and Influencing Factors in the Perspective of Green Economy--Taking Anhui Province as an Example."
22 RGGI in eight provinces and cities covering twelve northeastern states are the typical regional carbon market; South Korea, Kazakhstan, New Zealand, Norway, Switzerland and China national carbon trading system launched in 2017 all belong to the national carbon market; and carbon trading under Kyoto mechanism is a typical international market, covering more than 100 countries and regions, the most important is EU ETS covering 25 EU countries.

**Table 2.** Main forms of carbon market connectivity[23]

| Scope | | Target Quota | Project |
|---|---|---|---|
| International | 1 + 1 | EU-ETS + Norway ETS<br>EU-ETS + Swiss ETS<br>WCI | JI |
| | 1 + 0 | EU-ETS + AUS | CDM<br>California + Canada / Mexico |
| Chinese | 1 + 1 | — | CCER |
| | 1 + 0 | Beijing + Hebei + Inner Mongolia | CCER |

*3.3.2 Micromarket structure*

The micromarket structure includes three categories: primary market and secondary market, exchange trading market and over-the-counter market, and financing service market and support service market.

First, the significance of the distinction between primary and secondary markets lies in the following: the primary market is the issuing market, and the secondary market is the trading market.[24]

---

23 Note: 1 connection between the two carbon pricing zones; 2 links between the carbon pricing zones and noncarbon pricing zones; 3 originally scheduled to start in 2016 but is currently in negotiation; 4 originally scheduled to start on 1 July 2015, but Australia abolished the carbon emission trading system proposed in 2015 in July 2014.

24 In carbon financial market, the primary market is to create carbon emission quotas and project emissions of two types of basic carbon assets (carbon credit) market, the generation of carbon quota mainly through free distribution and two way auction, the project emissions are according to the corresponding methodology to complete project certification, monitoring, project for the record and emissions and a series of complex procedures, when the carbon quota or project emissions completed in the registration register, became its holding institutions and can formally conduct trading, performance and use of carbon assets. Secondary market is carbon spot and carbon financial derivatives trading circulation market, is the hub of the carbon financial market; despite being on the outside, the secondary market can, gather all kinds of assets, help participants find counterparties, finding price, complete silver delivery settlement through market main body; the secondary market can also improve market

The primary market is the foundation of the secondary market, and the type and quantity of carbon assets put in the primary market directly determine the scale and structure of spot circulation in the secondary market.

Second, the differentiation of the exchange market and the over-the-counter market lies in that the secondary market includes two parts: the floor exchange market and the over-the-counter market.[25] At present, floor trading is the mainstream in the global carbon financial market, but over-the-counter trading still occupies an important position. After the global financial crisis in 2008, most over-the-counter trading began to turn to the floor for clearing to avoid trading risks.

Finally, the significance of the distinction between the financing service market and support service market lies in the following: the key to carbon financing services is reasonable pricing and valuation, and the other is to determine the realization ratio, which directly determines the financing scale and risk distribution.[26] The products and services such as carbon insurance and carbon index that support the service market not only insure and increase credit for the carbon assets in the financing service market but also provide information guidance for the secondary trading market and can become the technical basis for the indexed trading tools of the secondary market.

---

liquidity, provide participants with hedge risk and hedging way through the introduction of all kinds of carbon financial trading products and services.

25  Exchange trading refers to the trading of carbon assets in centralized trading places (approved exchanges). This kind of trading has fixed trading places and trading time, and open and transparent trading rules. It is a standardized and organized trading form, and the trading price is mainly determined through bidding and other ways. Over-the-counter trading, also known as over-the-counter trading, refers to various carbon asset trading activities conducted outside the exchange through nonbidding, and the price is determined by the two parties through negotiation.

26  The financing service market is often closely related to the secondary trading market, and the transaction price of the secondary market can often become the acceptable pricing and valuation basis for both parties other than the third-party evaluation. At the same time, when financing service default occurs, the carbon assets can also be handled conveniently and quickly by relying on the secondary market. Carbon financing service market through carbon assets pledge, carbon asset repurchase, carbon assets custody, and other products and services, for carbon assets provide a low risk of value; carbon asset owners who opened up a new financing channels will control the row of precipitation in the register of most carbon quota assets activation and circulation, at the same time the future usufruct of discount project business, is of great significance to the carbon asset management.

**4 The current state of the carbon financial legal system**

The carbon financial legal state of the system includes four faces: the environmental protection legal system, the carbon financial quota initial allocation legal system, the carbon financial trading legal system and the carbon financial regulatory legal system. Carbon reduction collaborative control is the root of the regulation of carbon finance, but the initial allocation of carbon emissions and the carbon trading market laid a solid foundation for carbon financial trading and will command the control of environmental protection and market leading market trading and market regulation law, thus forming the carbon emissions financial instruments properties. This provides financial support for the synergistic mechanism of carbon reduction.

*4.1 The environmental protection legal system dimension*

Response to climate change is a major issue of environmental protection, and some scholars have noted that the climate change legal framework should be "two wings" architecture, namely, the basis of which is building the basic legal system of climate change, through mitigation legislation and adaptation legislation to climate change, and thus putting the carbon financial system into the slow legislative framework.[27] From the perspective of environmental law, the environmental protection law that is most related to carbon finance is the Air Pollution Prevention and Control Law. Previous results in theory and practice have proven that the coordinated control of air pollutants and greenhouse gases can bring synergistic effects. The law of the prevention and control of atmospheric pollution in China article 2 is as follows: prevention and control of atmospheric pollution, should strengthen the coal, industrial, motor vehicles, dust, agriculture and other comprehensive control of atmospheric pollution, in addition to implementation of regional joint prevention and control of atmospheric pollution, particulate matter, sulfur dioxide, nitrogen oxides, volatile organic compounds, ammonia and other atmospheric pollutants and coordinated control of greenhouse gases.[28] This clause reflects the legal layout of "coordinated control of air pollutants and greenhouse gases", and this is the principle provision of the Prevention and Control Law of Air Pollution.[29] This clause is a preliminary attempt to bring greenhouse gas emission reduction into the legal framework control, to change the current state of legal absence that cannot be followed for greenhouse gas emission reduction, and to lead China to take the initiative in the new international order against climate change.[30] "Collaborative control" refers to the inclusion of greenhouse gas emission reduction into the framework of the Air Pollution Prevention and Control Law based on the correlation characteristics between air pollutants and greenhouse gases to obtain multiple synergistic effects of simultaneous emission reduction, system cost reduction and management cost reduction.[31] "Collaborative control", as the legal basis and the main way of controlling greenhouse gas emission reduction, includes three aspects: subject, object and content.[32]

---
27 Zhang Z (2010) A preliminary study on the framework system of China's climate change response law. J Nanjing Univ (Humanit Soc Sci Ed).

However, this perspective is not sufficient to explain the development of carbon finance.[33] The

---

28  Liu J (2018) Legal path of greenhouse gas emission reduction: collaborative control of greenhouse gases and air pollutants —— review article 2, paragraph 2 of the air pollution prevention and control law. J Xinjiang Univ (Philos Humanit Soc Sci Ed).
29  Cao M, Cheng Y (2015) My opinion on the revision of the air pollution prevention and control law: on the air pollution prevention and control law (revised draft). Jianghuai Forum.
30  Zhou X, Zhang J (2017) Legal thoughts on the collaborative governance of air pollution and climate change. Soc Sci Forum.
31  Yao Y (2014) An analysis on the regulation of greenhouse gas emission. Environ Prot.
32  Coping climate change is a multilevel and multilatitude complex and a comprehensive system project, involving greenhouse gas emission reduction, energy conservation, energy efficiency, renewable energy utilization, carbon emission trading, carbon sink and other fields. Both the greenhouse gas emission reduction legislation and the carbon emission right trading legislation belong to the "end-control" legislation. In the legal framework of climate change, the content of collaborative control is mainly reflected in the process of greenhouse gas emission reduction and carbon emission trading with air pollution prevention and control. Regarding the carbon emissions trading in this field, because our country is responding to climate change in the construction of the rule of law, not a single legislation in terms of commanding the global situation mobilizes the parties to deal with comprehensive climate change or law framework, but in the field of climate change, through decentralized special law, such as making the interim measures for carbon emissions trading management, or adding or modifying specific terms, the conditions mature after a unified climate change law , including carbon emissions trading.
33  In March 2018, the National People's Congress passed a new round of reform of party and state institutions, vigorously promoted the ecological civilization construction in the field of the integration of ecological environment, formed the original functions of the environmental protection department (including prevention and control of atmospheric pollution) together with profound reason behind the development of carbon finance is to change the economic development mode, build an environmentally friendly government and develop green financial institutions.[34] Therefore, although environmental protection laws are the reason for the emergence of carbon finance, that is, the control of carbon emissions, they are not enough to cover and fully regulate the field of carbon finance. Therefore, the regulation of carbon finance needs to establish a perfect special legal framework, including three aspects: initial distribution, financial trading and financial supervision. The following are discussed from three perspectives: initial allocation of carbon financial quota, carbon financial trading system and carbon financial supervision system.

*4.2 Carbon financial quota initial allocation of the legal system orientation*

The legal system of the initial allocation of carbon financial quotas is essentially the legal system of the quota allocation of carbon emission rights. The development of China's carbon emission rights market has gone through a process from local pilot to gradually becoming a standard to the opening of the national market. Since 2021, China's carbon emissions trading ETS national market regulatory legislation has taken effect on February 1,2021 in terms of the measures for the administration of carbon emissions trading (trial) ", the ecological environment released in May 2021, and the carbon right market registration, trading and settlement rules, became ready for the national open market in line with the

---

the original principles of the National Development and Reform Commission to climate change and emissions responsibility unity to ecological protection, respectively set up the atmospheric environment and climate change being responsible for the specific work. The adjustment "is a major institutional arrangement, whose core goal is to achieve the synergistic benefits of reducing atmospheric conventional pollutants and greenhouse gas emission control". Therefore, now that the background of the supervision and management system has changed greatly, and the ecological and environmental protection departments, as the appropriate subjects, have the right to jointly control the emission of greenhouse gases and air pollutants.

34 Coordination is the interaction between 2 active parties and the active party, with one being the dominant active party. At present, the rule of law for air pollution prevention and control is relatively sound, and the legislation on climate change is in the development stage. In the coordination relationship, the prevention and control of air pollution should be the active side and the receiver, and the climate change needs to effectively control air pollution, to achieve a win–win situation.

corresponding legislation.[35] At the same time, the Shanghai Energy Exchange recently issued a notice clarifying the trading venues, trading methods, trading periods, trading accounts and other matters,[36] further clarifying the direction for the construction of the carbon emission rights market and laying a solid foundation for the development of the carbon financial market.[37] Based on past experience, the setting method of the total carbon emission quota in China is generally divided into three steps: the total carbon emission quota system, the total carbon emission quota system and the specific carbon emission quota, including the total quota allocated to the enterprise and the total quota reserved by the government. The total amount of quotas allocated to enterprises is the sum of quotas allocated to enterprises by the government according to the specific quota allocation plan. At present, China's local pilot carbon emission rights quota allocation system is still mainly free distribution. The carbon emission enterprises obtain further free quotas, with less gap, the purchase demand is mainly concentrated in the power industry, and most of the industry enterprises can complete the contract without buying quotas.[38]

Looking from the scope of carbon emission obligations, local carbon emissions trading pilots determine the emissions management main body on the legislation mode of the combination of the upper law into the subject of the emission management type, clear emissions scale of enterprises into the management scope, put in methods according to the provisions of the upper law and the specific emissions within the administrative area, and determine the specific enterprise scope and list. Specifically, according to article 8 of the Administrative Measures for Carbon Emission Trading (Trial), it can be seen that the obligation subject of carbon emission rights in China is the key greenhouse gas emission units that meet the above two conditions. In addition, China's various local pilot projects have carried out quota auctions. The target of the auction is mainly for the performance of the contract and active market liquidity, and the overall proportion of the auction quota is not high.[39]

---

35 Liu C (2021a) Global law firm: industrial investment | Market overview and regulatory framework for carbon rights investment. https://mp.weixin.qq.com/s/_1Uua-5qXk3VjJnEmn_W8w.
36 Chen S (2011) Marginal cost reduction and China's environmental tax reform. China Soc Sci.
37 Shanghai Environment and Energy Exchange (2021) [Exchange announcement] announcement on matters related to national carbon emissions trading. https://mp.weixin.qq.com/s/fFA8HSfZ4qzQb2nx0pO6Dg
38 Refer to the Implementation Plan for Setting and Allocation of National Carbon Emission Trading Alotas in 2019-2020 (Power Generation Industry) (Draft for Comments).

*4.3 Carbon financial trading legal system orientation*

The essence of the legal system of carbon financial trading lies in the financial trading contract system, realizing capital financing through continuous market trading and providing a steady stream of capital supply for carbon financial trading. According to Article 9 and Article 509 of the Civil Code, the performance of contracts should avoid the waste of resources, pollution of the environment and destruction of the ecology.[40] The carbon financial market includes the primary market and secondary market, so carbon financial activities in financial transactions, namely, various financial investments and capital operations, for the waste of resources, environmental pollution and ecological destruction should be accordingly reflected in responsibility, and more specifically embodied in the rights and obligations agreed upon in the financial transactions and the contract.[41] Therefore, analyzing the current state of carbon finance-related contracts is helpful for understanding the legal system of carbon finance trading.[42]

---

39 Greenhouse gas emission units that meet the following conditions shall be included in the list of key greenhouse gas emission units (hereinafter referred to as key emission units): (1) belong to the industries covered by the national carbon emission trading market; (2) the annual greenhouse gas emissions reach 26,000 tons of carbon dioxide equivalent.
40 That is, the implementation of the green principle in the performance of the contract is to avoid the waste of resources, pollution of the environment and damage to the ecology, and to avoid the waste of resources, pollution of the environment and ecological destruction as an obligation of the parties to the contract. However, the law does not stipulate what kind of legal consequences the party should bear in violation of the obligation.
41 Carbon emission quota repurchase financing transaction refers to the transaction behavior in which the seller sells the carbon emission quota to the buyer at an agreed price and agrees to purchase the carbon emission quota from the buyer at another agreed price on a certain date in the future. Carbon emission quota repurchase financing transactions include initial transactions and repurchase transactions. Accordingly, the carbon emission quota repurchase financing contract is a contract for the rights and obligations of the buyer and the seller and related matters agreed upon for such transactions.
42 The carbon emission quota pledge contract is a written guarantee contract in which the pledgor and the pledgee guarantee the performance of the carbon emission quota. The pledgor shall hold the carbon emission quota to the pledgee and set the pledge guarantee to the pledgee. When the debtor of the principal contract fails to perform the principal debt to the pledgee, the pledgee may have the priority of the proceeds from the carbon emission quota according to law.

At present, carbon financial trading contracts mainly include carbon emission quota repurchase financing contracts, carbon emission quota pledge contracts, over-the-otc option trading contracts of carbon emission rights,[43] and over-the-otc swap trading contracts of carbon emission rights.[44]

*4.4 Carbon financial regulatory legal system orientation*

At present, although the carbon trading pilot only allows the spot trading of carbon emission quotas, carbon financial products based on carbon emission rights quotas are an important development direction in the future. The development of carbon financial products is conducive to providing more liquidity for carbon emission trading. China is currently actively exploring the innovation of carbon financial products at the national level and at the local level.(see table 3)[45] China's current policies related

---

43  Over-the-counter option of carbon emission right is an oversite nonstandardized carbon financial innovation product entrusted by the trading parties with the carbon emission right (EUA or CCER) and the exchange to supervise the execution of the EUA or CCER. Both parties shall determine the exercise period and the execution price upon the signing of the contract, and after the option buyer makes the decision of execution or nonexecution during the exercise period, they shall entrust the exchange to complete the contract execution work according to the agreement of both parties. The OTC option trading contract of such carbon emission right is concluded by the two parties and the exchange.
 In addition, some carbon emission option trading contracts are only signed by both parties, stipulating that the carbon emission right shall be the subject matter. One party has the right to purchase the agreed subject matter from the other party at the agreed exercise price within a certain period of time in the future, and pay the agreed contract price (premium) to the other party. Under this trading arrangement, the exchange does not join the option trading contract of both parties, and the parties can go through the relevant registration procedures of purchasing the carbon emission right at the exercise to the agreed exchange. Of course, if for the agreed exercise period, the right party decides not the right, there is no need to have a relationship with the exchange.
44  The over-the-counter swap transaction of carbon emission right is an over-the-counter contract transaction with the difference between fixed price trading and floating price trading. The parties to the transaction determine the transaction at a fixed price at the time of signing the contract, and agree in the contract to complete the reverse transaction corresponding to the fixed price transaction at some future time at the current market price. In the final settlement, the two parties only need to cash settle the difference between the prices of the two transactions. The over-the-counter swap trading contract of carbon emission right is concluded by the two parties and the exchange otc swap trading provides a means for carbon market trading participants to hedge price risks and carry out hedging over the counter. At the same time, it can indirectly create liquidity for enterprises to manage carbon assets, and this is an important innovation in the field of carbon finance.
45  In June 2012, the National Development and Reform Commission promulgated the Interim Measures for the Management of Voluntary Greenhouse Gas Emission Reduction Transactions, defining and standardizing the trading market of China Certified Voluntary Emission Reduction (CCER) projects in terms of trading products, trading subjects, trading places and trading rules, registration and regulatory system, the National Development and Reform Commission issued the Supporting Greenhouse Gas Voluntary Emission Reduction Projects, and clarified the filing requirements, working procedures and reporting format of voluntary emission reduction projects. In December 2014, the National Development and Reform Commission issued the Interim Measures for the Management of Carbon Emission Trading, setting up the basic framework of the national unified carbon emission quota and trading market, and systematically standardizing its development direction, ideas, organizational structure and the design of related basic elements. In terms of the development of carbon finance, the ecological environment on February 1,2021 of the carbon emissions trading management measures (trial) " monk in the opinions of the provisional regulations on carbon emissions trading related legal system of the preliminary construction, among them, the latest legislative purpose of the provisional regulations is to regulate carbon emissions trading, strengthen the control and management of greenhouse gas emissions, promote carbon dioxide emissions peak and carbon neutral vision, promote economic and social development to green low carbon transformation, promote the construction of ecological civilization. The introduction of legal and policy documents to climate change and carbon finance are shown below.

---

related to carbon emission trading has brought about several changes, including the establishment of the national, provincial and municipal carbon emission trading supervision system for the first time, strengthening the responsibility and rights status of enterprises in carbon emission trading, and refining carbon emission trading to county-level units. In addition, in 2014, China Securities Regulatory Commission said that it would continue to innovate the varieties of the futures market, promote the pilot work of carbon emission right trading, and study the feasibility of carrying out domestic carbon futures trading. In April 2015, The State Council issued the General Plan of China (Guangdong) Pilot Free Trade Zone, proposing to establish an innovative futures exchange with carbon emission rights as the first type in the Guangdong Free Trade Zone. In January 2016, the CSRC mentioned again at the National Securities and Futures Regulation Work Conference the need to study and demonstrate carbon emission right futures trading, and explore the use of market-oriented mechanism to promote green development. On August 31,2016, the People's Bank of China, Ministry of Finance, National Development and Reform Commission, Ministry of Environmental Protection, China Banking Regulatory Commission, Securities Regulatory Commission and CIRC jointly issued the guidance on constructing green financial system to develop various carbon financial products, promote the unified carbon emission trading market, carbon forward, carbon swap, carbon option, market, and explore carbon emission right futures trading.

**Table 3.** Review of policies related to carbon finance

| Policies | Date of issue | Policy document/important statement | Relevant content |
|---|---|---|---|
| Carbon constraint target | | | |
| | June, 2007 | China's National Plan on Climate Change | China has identified specific goals, basic principles, key areas and specific measures to address climate change by 2010. |
| | August, 2010 | On the Pilot Work of Low-carbon Provinces and Low-carbon Cities | We will study the use of market mechanisms to achieve emission reduction targets. |
| | December, 2011 | Work Plan for Controlling Greenhouse Gas Emissions during the 12th Five-Year Plan Period | The overall requirements and main objectives of the control and discharge by 2015 are clarified. |
| | October, 2016 | Work Plan for Controlling Greenhouse Gas Emissions during the 13th Five-Year Plan Period | The overall requirements and main targets for emission control by 2020 are clarified, and proposed to promote China's carbon dioxide emissions to peak in approximately 2030 and strive to reach the peak as soon as possible. |
| Mechanism planning | | | |
| | October, 2021 | Opinions of the CPC Central Committee and The State Council on Complete, Accurate and Comprehensive Implementation of the New Development Concept and achieving carbon Peak and Carbon Neutralization | The overall requirements and main targets by 2025, 2030 and 2060 have been defined, and carbon peak and carbon neutral into overall economic and social development to ensure that carbon peak and carbon neutral as scheduled. |
| | December, 2021 | Comprehensive Work Plan for Energy Conservation and Emission Reduction during the 14th Five-Year Plan Period | The overall requirements and main targets by 2025 have been clarified to achieve synergies in energy conservation, carbon reduction, and lay a solid foundation for achieving the goal of carbon peak and carbon neutrality. |
| Emission reduction commitment | November, 2009 | The Copenhagen Climate Conference | In 2020, carbon emissions per unit of GDP will be reduced by 40-45 percent compared with 2005 levels, and a unified national statistics, monitoring and assessment system will be established. |
| | November, | The Sino-US Joint Statement on Climate | Carbon emissions will peak in approximately 2030 and efforts |

| | 2014 | Change | will peak as soon as possible, and the proportion of nonfossil energy in primary energy consumption will increase to 20%. |
| --- | --- | --- | --- |
| | December, 2015 | 《Paris Agreement》 | Control the global average temperature rise in the preindustrial 2º C level and try to reach 1.5º C below. |
| | September, 2015 | China–US dollar Joint Statement on Climate Change | Carbon emission intensity per unit of GDP will be reduced in 2030 by 60-65 percent compared with 2005 levels. |
| | September, 2020 | General Debate of the 75th United Nations General Assembly | Carbon dioxide emissions aim to peak by 2030 and strive to be carbon neutral by 2060. |
| | November, 2021 | The Glasgow Agreement, | Approved the Paris Agreement goal of "keeping temperature rises within 1.5 C"; pledged to cut global carbon dioxide emissions by nearly half by 2030. |
| | November, 2021 | The Glasgow Joint Declaration on Strengthening Climate Action in the 2020s | Intensive climate action under the Framework of the Paris Agreement in the 2020s to enable the goal of temperature rise limits to be achieved and to work together to identify and address relevant challenges and opportunities. |
| Carbon market policy Central policy declaration | September, 2015 | The Overall Plan for reforming the System for Promoting Ecological Civilization | We will deepen trials for carbon emission trading and gradually establish a national market for carbon emission trading. |
| | September, 2010 | The Decision on Accelerating the Cultivation and Development of Strategic Emerging Industries | We will establish and improve trading systems for major pollutants and carbon emissions. |
| | October, 2010 | During The 12th Five-Year Plan period | We will significantly reduce energy consumption intensity and carbon emission intensity as binding targets, and gradually establish a carbon emission trading market. |
| | December, 2012 | Report to the 18th National Congress of the CPC | We will actively carry out trials of carbon emission rights trading. |
| | November, 2013 | Resolution of the Third Plenary Session of the 18th CPC Central Committee | Further clear requirements, the implementation of carbon emission rights trading system. |
| | September, 2015 | China–US dollar Joint Statement on Climate Change | In 2017, the national carbon emission trading system was launched, covering six key industries: steel, electric power, |
| | | | chemical industry, building materials, paper making and nonferrous metals. |
| | October, 2015 | Resolution of the Fifth Plenary Session of the 18th CPC Central Committee | We will establish a sound initial distribution system for energy, water, sewage and carbon emission rights. |
| | October, 2017 | The report to the 19th CPC National Congress | We will establish a market-based and diversified ecological compensation mechanism. |
| | October, 2021 | Opinions of the CPC Central Committee and The State Council on Complete, Accurate and Comprehensive Implementation of the New Development Concept and achieving carbon Peak and Carbon Neutralization | The overall requirements and main targets by 2025, 2030 and 2060 have been defined, and carbon peak and carbon neutral into overall economic and social development to ensure that carbon peak and carbon neutral as scheduled. |
| | October, 2022 | The "Twenty-ten reports" | We will actively yet prudently promote carbon peak and carbon neutrality, and improve the market trading system for carbon emission rights. |
| The carbon market is responsible for the department's work deployment | October, 2011 | Notice on Developing the Pilot Work of Carbon Emission Right Trading | Approved the seven provinces and cities of Beijing, Tianjin, Shanghai, Chongqing, Guangdong, Hubei and Shenzhen to carry out the pilot carbon emission right trading in 2013. |
| | June, 2012 | Interim Measures for the Administration of Voluntary Greenhouse Gas Emission Reduction Trading | The CCER project development, transaction and management are systematically standardized. |
| | 2012 | | The first batch of national voluntary emission reduction trading institutions, including the Beijing Environmental Exchange. |
| | October, 2012 | Guidelines for the Approval and Certification of Voluntary Greenhouse Gas Emission Reduction Projects | The filing requirements, working procedures and reporting formats of the CCER project approval and certification institutions are stipulated. |
| | December, 2014 | Interim Measures for the Administration of Carbon Emission Rights Trading | Normative requirements are put forward for the development direction, thinking, organizational structure and the design of related basic elements of the national unified carbon emission right trading market. |
| | January, | Notice on Effectively Starting the Key Work of | Eight industries participating in the national carbon market |

| | | |
|---|---|---|
| 2016 | the National Carbon Emission Trading Market | have been identified, and the historical carbon emissions of enterprises to be included should be MRI, and the accounting reports of the supplementary carbon emissions of enterprises have been put forward. |
| December, 2020 | Administrative Measures for Carbon Emission Rights Trading | Normative requirements are put forward for the development direction, thinking, organizational structure and the design of related basic elements of the national unified carbon emission right trading market. |
| May, 2021 | Management Rules for Carbon Emission Right Registration (Trial), Management Rules for Carbon Emission Right Trading (Trial) and Management Rules for Carbon Emission Right Settlement (Trial) | Detailed normative requirements for the registration, trading and settlement of unified carbon emission rights. |
| July, 2021 | The national carbon market has officially launched its online trading. | The launch of the national carbon market's online trading marks a significant step towards formalizing carbon emissions trading in China. It enables efficient, transparent trading of carbon credits, helping to regulate emissions, promote sustainability, and contribute to achieving the country's carbon neutrality goals. |
| May, 2024 | The Provisional Regulations on the Management of Carbon Emission Trading | establish carbon trading at the national legal level, clarifying the rights, duties, and responsibilities of participants. This legal framework ensures a structured, transparent carbon market, supporting China's carbon neutrality goals and aligning with global climate efforts. |

Therefore, according to the characteristics of carbon emission quota trading, the normal development and effective supervision of carbon finance can be separated into two aspects: the first aspect establishes the quota registration system to track the specific trading quota to ensure the safety of market trading and the legal supervision of the quota trading market to ensure the legitimacy of market trading.[46] Although a large number of pilot studies and practices are being carried out in the field of carbon finance, the construction of carbon finance systems still faces a large number of problems, and many institutional obstacles need to be urgently solved.

*5 Problem: the obstacles to the construction of a carbon financial system*

The development of a robust carbon financial system in China faces significant challenges. Key obstacles include the restrictive economic structure, the need for an improved monitoring, reporting, and verification (MRV) system, and inefficiencies in the implementation of local pilot projects. Despite the establishment of national carbon trading platforms, issues such as limited market liquidity, lack of diverse financial products, and insufficient regulatory frameworks continue to hinder market activity. Additionally, problems like inadequate information disclosure and the absence of a carbon financial judicial guarantee system further constrain the growth and stability of the carbon finance market.

*5.1 The restrictive factors of the economic structure*

At present, traditional coal and oil are still the main sources of energy consumption in China.[47] The reason is that the pillar industry of the national economy lies in the high energy consumption industry, and the growth of GDP and the improvement of national living standards are based on this. In the short

---

term, the industrial structure should be orderly, and there are still great difficulties in eliminating backward production capacity and accelerating structural adjustment. On the one hand, it involves the interest connection and shelter of heavily polluting enterprises and local governments. On the other hand, China is still in the development period of industrialization and urbanization, objectively determining the difficulties of structural adjustment. The difficulty facing the development of a low-carbon economy lies in that it only hears the voice and does not see the person. The excessive supply-side structure at the economic level leads to the limited development space of carbon finance.

*5.2 Inadequate construction of the carbon financial market*

The inadequate construction of the carbon financial market in China presents several challenges that hinder its effective functioning. Key issues include the need for an improved monitoring, reporting, and verification (MRV) system, where the credibility of third-party verification and transparency are essential for ensuring reliable carbon data. Despite the launch of the national carbon trading platform, market activity and liquidity remain low, exacerbating issues in carbon emissions control. Moreover, the market's single carbon trading product and limited auction participation restrict development, and a lack of liquidity further impedes carbon trading. Additionally, the carbon emissions information disclosure and legal frameworks are still underdeveloped, affecting market transparency and investor confidence.

*5.2.1 The monitoring, reporting and verification (MRV) system needs to be improved*

The qualification and independence of the third-party verification structure in the MRV system affect the transparency and credibility of government supervision.[48] The specific obstacles lie in the cumbersome approval procedures and lack of operability, easily leading to power rent-seeking; such obstacles are manifested in the imperfect regulatory system, the imperfect qualification examination system, reporting system, trading behavior registration system and information disclosure system. Although the national carbon trading platform was launched in Shanghai at the end of June, the subsequent activity and liquidity problems of the carbon financial trading market are still greatly being adjusted, so the construction of the MRV system is crucial. From the perspective of the local pilot projects, the localization requirements of the registration place of verification institutions and the professional level of inspectors have an important impact on the independence of verification institutions, the order of competition and the accuracy of carbon data.

---

48 The MRV system mainly includes monitoring (M easure), reporting (R eport), and verification (V erification). Monitoring is the collection process of carbon emission data and information, the report is the process of data submission or information disclosure, and the verification is the regular review or third-party assessment of carbon emission reports. These three elements are important foundations and guarantees to ensure the accuracy and reliability of carbon emission data.

*5.2.2 There are still great problems in the implementation effect of local pilot projects*

The original intention of the construction of the carbon finance market was to control the total carbon emissions and achieve the purpose of energy conservation and emission reduction.[49] However, in recent years, after the pilot project, the total control effect is not obvious, which is a large gap compared with the emission reduction effect of the carbon finance market.[50] Currently, in the local pilot, total quota allocation is lax, the free auction proportion is too low, the emissions subject scope is too narrow, there is a lack of carbon trading products, seriously restricting the development in the field of carbon finance, and the national carbon trading platform after opening is still not enough to solve the problem of endogenous power of carbon financial development, which will greatly restrict the carbon financial market trading activity.[51]

---

49  Song X (2021) BinHui: from local carbon pilot to national carbon trading center. China Environ Manag.
50  At present, the liquidity of carbon trading market in China is not strong, the main for this reason is that the trading product is single and it is only spot.
51  See Zhang L (2020) A perspective on the development of China's carbon market. China Finance.

*5.2.3 The allocation methods and methods of carbon quotas need to be optimized*

At present, the quota allocation method is a relatively single, mainly free distribution. Although there are some auctions in some places, the proportion is not high and it fails to play a large role.[52] At present, carbon quota allocation in China mainly adopts the base method and the historical method. From the pilot experience card, the benchmark method to calculate carbon quotas requires high methodology, following the principle of unified industry allocation standards; a carbon quota that is too high is not conducive to achieving the effect of emission reduction, and a carbon quota that is too low will damage the enthusiasm of enterprises. In addition, the price difference of the paid allocation of carbon quotas among the regions in the eight pilot markets is bound to aggravate the disorderly competition among the regions and affect market fairness.

5.2.3.1 Lack of types of carbon financial products

Carbon financial products are the key to the development of the carbon financial market. As seen from the previous analysis of the current state, the most active carbon financial products in the current market are mainly carbon asset pledge, carbon fund and quota custody business. Even in the scope of the whole carbon trading market, there are only two carbon trading products: project type and quota type, with few carbon trading products and a single market product type. It does not play the function of market hedging and risk avoidance.[53] More importantly, the support for carbon financial innovation remains weak. From the perspective of local pilot projects, the low degree of market activity, the connection between regional quotas and national quotas, the lack of carbon financial product regulatory regulations, and the lack of carbon financial knowledge of emission control enterprises and financial institutions

---

52 Song X (2021) BinHui: from local carbon pilot to national carbon trading center. China Environ Manag.
53 Yang J (2018) The development status, problems and countermeasures of China's carbon trading market. Enterp Econ.

hinder the development of carbon financial products.[54]

5.2.3.2 Enterprises are not strong enough to use carbon finance

Enterprises are generally reluctant to use carbon financial instruments due to the lack of liquidity in the carbon financial trading market and the imperfect disclosure mechanism of carbon emission information.

5.2.3.3 Lack of liquidity in the carbon financial trading market and inactive trading

The main reasons for the lack of liquidity in the carbon trading market are the small market size, the single product variety, and the reluctance of emission control enterprises to take risks for carbon trading. China's carbon financial market trading has not yet formed a complete trading price mechanism, and coupled with the restrictions of the trading mode, the carbon price fluctuates greatly. Taking the performance period as the node, the carbon price and trading volume fluctuate greatly; the carbon trading volume during the nonperformance period is relatively small, highlighting the lack of market liquidity. In addition, the carbon financial market is less active and transparent, making it difficult to attract financial institutional investors to carry out transactions, increasing the risk of market manipulation.

5.2.3.4 The carbon emission information disclosure mechanism needs to be improved

At present, the development of carbon emission information disclosure in China is in the primary stage, and the lack of a reasonable pricing mechanism leads to the price of carbon emission trading rights being largely disturbed by human factors. The real value of carbon emission rights trading has not yet been grasped, and the market vitality is insufficient.[55] As a result, there is a carbon emission information disclosure system with some problems, such as the lack of unified disclosure rules and standards in the carbon financial market, lack of a carbon data audit system, lack of motivation to disclose carbon emission information, and concerns about the confidentiality of business information. Although the Ministry of Ecology and Environment has issued a notice on the issuance of the environmental information disclosure system, requiring enhanced environmental and social governance, more detailed rules still need to be further discussed and formulated.

---

54 Chen S, Li Z (2020) An introduction to green finance. Fudan University Press.
55 Yang J (2018) The development status, problems and countermeasures of China's carbon trading market. Enterp Econ.

5.2.3.5 The carbon financial judicial guarantee system has not yet been formed

Through case database retrieval and research of carbon financial market-related judicial cases we found that the current 14 cases only involve carbon emissions trading, including one involving more complex carbon emissions trading, namely, (2020) 01 people end 23215 cases. The dispute occurred in the Guangzhou carbon exchange, hereinafter referred to as the Guangzhou micro carbon case.[56] The case is a dispute between an investment institution and a carbon emissions trading platform. The investment institutions took the carbon emissions trading platform to the court with the notion that the trading platform should verify and ensure that its counterparty funds are in place. In the case of investment institutions that have delivered carbon quotas, the trading platform does not receive its counterparty (control unit) deposit involved in transfer money and exemption and shall bear the corresponding legal responsibility. In the reasoning part of the court, the acceptance court mainly decided that the agency has no claim for the basis of the contract relativity, the contract agreement and the legal responsibility. There is a very serious problem: the court has no high-level law as the basis for carbon trading-related procedures and conditions, leading to unreliable auditing of both types of trading and leading to disputes.

---

56  Civil judgment of second instance on contract dispute between Microcarbon (Guangzhou) Low Carbon Technology Co., Ltd. and Guangzhou Carbon Emission Trading Center Co., Ltd.; Civil judgment of first instance on contract dispute between MicroCarbon (Guangzhou) Low Carbon Technology Co., Ltd. and Guangzhou Carbon Emission Trading Center Co., Ltd

The current situation is that there is no high-level legislation regulation, leading the court not to act in accordance with the relativity of the contract. At present, the legal provisions related to carbon emission trading are still imperfect, the existing legal level is limited, and the scope of the rights and responsibilities of the relevant auxiliary subjects in the carbon emission trading market is indeed vague or even blank. It fully exposes that the current judicial organs for carbon trading cases of institutional guarantee are weak. If the protection of carbon trading cases is so weak, then carbon financial cases of judicial protection are impossible to handle. In the future, with the development and maturity of the carbon financial system, carbon financial disputes will soon appear. Recently, the carbon financial judicial security system should be implemented for necessary improvement and development in response to the gradual expansion of the carbon trading market and carbon financial market.

*6 Suggestion: Countermeasures for carbon financial system construction in the future*

The future development of China's carbon financial system requires comprehensive legal countermeasures to address current challenges and ensure its success. Key actions include raising the legislative level of carbon quota allocation to provide a solid legal foundation, improving judicial interpretations to guide carbon-quota trading, and enhancing environmental information disclosure mechanisms through the incorporation of Environmental, Social, and Governance (ESG) standards. Additionally, promoting a specialized environmental judicial system will help address the growing number of carbon-related disputes, ensuring legal clarity and stability. These legal reforms are crucial to fostering a well-functioning carbon financial market and supporting the achievement of China's carbon neutrality goals.

*6.1 Carbon, financial and economic countermeasures*

The construction of a robust carbon financial system is pivotal for advancing China's low-carbon transition and achieving carbon neutrality targets. A key countermeasure involves gradually adjusting the economic structure to incorporate carbon pricing, which helps align financial markets with climate goals. By introducing carbon financial products such as carbon futures and carbon bonds, the market can better manage risks and promote sustainable industries. This adjustment process also facilitates energy conservation and emission reduction while reshaping industrial structures to support low-carbon development. Financial institutions, alongside government policy, must innovate to diversify carbon financial products and strengthen market infrastructure to stimulate carbon finance activity.

*6.1.1 We will gradually adjust the structure of economic development*

Carbon trading plays a fundamental role in the market mechanisms to address climate change, enabling carbon prices to reflect the degree of resource scarcity and the cost of pollution control. The carbon market trading scale of continuous expansion greatly promoted the monetization of carbon and prompted carbon emissions to gradually derive into the rapid flow of financial assets—various carbon financial products. The carbon finance system is essential for facilitating the transition to a low-carbon economy through transition finance. This system supports the financing of projects that reduce carbon emissions while promoting sustainable development. Transition finance involves directing funds to industries and sectors that are undergoing a transformation towards greener practices, such as clean energy and sustainable manufacturing. By incorporating legal frameworks, the system ensures that investments align with climate goals, fostering a structured and accountable approach to mitigating climate change and achieving carbon neutrality targets.[57] Its price discovery function can lead market participants to make more reasonable estimates of carbon trading product prices. Market constraints under the equilibrium price will lead investors in the carbon market to develop a more effective trading strategy and risk management decisions. Various carbon financial products include carbon futures, carbon options and carbon asset securitization. With the vigorous development of China's carbon financial market, more carbon financial derivatives are bound to emerge in the future. Optimizing the allocation of resources through the carbon price elements of the carbon market is required, along with using the internal market mechanism to drive energy conservation and emission reduction, greatly reducing the emission reduction cost, and controlling the development of industries with high energy consumption and high pollution. Through financial innovation, the introduction of new carbon financial products can change the passive state of carbon trading in China and promote the development of energy conservation and emission reduction work and industrial structure adjustment.[58] By advocating the green development concept, enterprises should strengthen their consciousness of energy conservation and emissions reduction, adopt more varieties of carbon financial products, gradually promote carbon financial market trading activity, form financing, more money will flow into low carbon economic development, further improving the economic structure of development, eliminating backward production capacity, and forming ecological mutual promotion of a good pattern of civilization and economic development. Fig. 1 takes carbon funds as an example to explain the mechanism of carbon financial adjustment and the influence of economic and industrial structure.

---

57  Pan, Xiaobin, and Liu Shangwen. "Exploring the Pathways of Legalizing Transition Finance in China in the Context of Climate Change." Advances in Climate Change Research 20.4 (2024): 465.
58  Lin M (2021) A study on the impact of carbon finance on China's industrial structure. Shanghai Bus.36

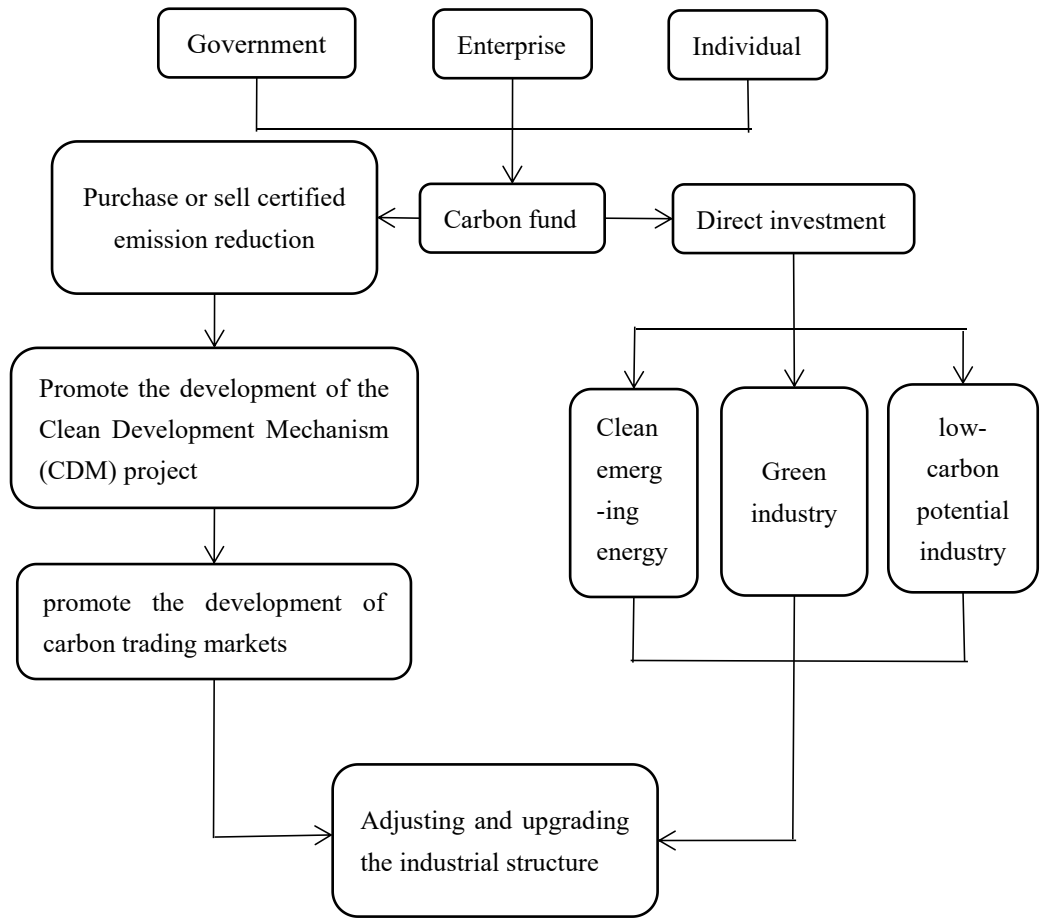

**Fig. 1** Impact mechanism of carbon funds on industrial structure

*6.1.2 We will strengthen the infrastructure construction of carbon financial trading markets*

The key to strengthening the infrastructure construction of the carbon financial trading market lies in building a multilevel and multiproduct carbon financial trading market. From the research experience of foreign scholars, improving the carbon pricing mechanism and strengthening the prudential supervision of carbon finance are the key means to strengthen the infrastructure of the carbon finance market. Emanuele Campiglio noted that it is believed that carbon emission pricing is the prerequisite for the construction of a carbon financial system. However, due to market failure, carbon pricing alone is not enough. Moreover, it needs to play the role of monetary policy and financial macroprudential supervision on this basis to control the steady development of carbon finance with monetary policy tools.[59] Limei Sun Think can regulate the carbon emissions regulations of enterprise number, standardize the number of market intermediaries, establish market regulation level, develop carbon intensity reduction level to promote carbon financial tool innovation, and develop of low carbon economy; at the same time, the government can formulate new carbon market related policies and regulations, influence the relationship between the carbon market and the main body in a timely manner, and promote the development of low carbon technology.[60] In the future, strengthening the research on carbon financial products, strengthening infrastructure construction, and improving the construction of a national carbon trading platform based on Shanghai are the top priorities of the orderly development of carbon financial trading.

*6.1.3 Financial institutions are encouraged to develop and pilot more carbon financial products*

At present, China's carbon financial products have a single type, and this cannot meet the realistic needs of risk hedging and hedging of emission reduction enterprises. Financial institutions should be encouraged to continuously innovate and pilot carbon financial products, activate the market, and meet the multilevel and diversified investment and financing needs of customers.[61] The government should introduce relevant policies to encourage financial institutions to develop diversified carbon financial products, including carbon assets mortgage, carbon funds, carbon options, carbon bonds, carbon asset securitization, carbon trust and carbon insurance, and constantly enhance professional promotion with specialized carbon financial products, reduce the potential risk in carbon quota trading, reduce the

---

performance cost, revitalize the carbon financial assets, promote financing, and improve carbon financial services market influence.

*6.1.4 Strengthen the publicity and education of carbon finance*

China's current carbon financial market is still in the primary stage, and the public social awareness and participation in the carbon financial market is generally low, seriously influencing carbon financial in terms of social capital gathered. In the national carbon emissions trading platform completed in Shanghai, Shanghai must seize the valuable platform advantage, vigorously promote carbon finance, expand the social influence of the carbon financial market, and deepen the national carbon trading platform advantage. Shanghai was suggested through newspapers, television and network project awareness campaigns, organization society "low carbon star", "environmental economic star" selection and a variety of forms to strengthen carbon financial promotion, letting the public gradually know and accept carbon financial; awareness campaigns include public service ads, but can also use the mobile app more conveniently. Recently, alibaba "ant forest", Guangzhou carbon exchange development "Guangdong carbon trading", and Hubei carbon exchange development "carbon bag" have achieved good awareness effects.

## 6.2 Legal countermeasures of carbon finance

The legal framework surrounding carbon finance is crucial for ensuring the effective development of the carbon financial market in China. One of the primary legal countermeasures is raising the legislative level of carbon quota allocation. Currently, the existing regulations are departmental, and the legal status remains relatively low, hindering market activity. To address this, it is essential to accelerate the development of high-level legal systems, including the National Carbon Emission Trading Law and the Carbon Trading Law, which can provide the necessary legal backing for the carbon emission rights market. Strengthening these laws will ensure the stability and predictability of the market, fostering further growth and reducing uncertainties.

*6.2.1 Raise the legislative level of carbon quota allocation*

Although the Administrative Measures for Carbon Emission Trading (Trial) has been issued, the Interim Regulations on the Management of Carbon Emission Trading are still in the collection of opinions stage. That is, the current legislative level involving carbon emission trading is only departmental regulations, and the legislative level is low, and this is not conducive to the improvement of market activity. The carbon finance system involves various institutional mechanisms aimed at promoting sustainable development through financial instruments such as green credit and carbon credits. These mechanisms help direct investments into eco-friendly initiatives and technologies that reduce carbon emissions. By encouraging industries to adopt greener practices, the system supports both environmental sustainability and economic growth. Additionally, carbon finance frameworks help alleviate environmental and employment pressures, ensuring a balanced transition to low-carbon economies while fostering the development of sustainable business models and job creation opportunities.[62] In the future, we will accelerate the construction of high-level legal systems such as the Regulations on carbon emission trading, the Climate Change Law and the Carbon trading Law, build the operation of the national carbon market on a complete rule of law, and stabilize the institutional expectations and market expectations of trading entities.[63] At present, the most urgent task for the development of carbon finance is to issue the Interim Regulations on the Management of Carbon Emission Trading, the National Carbon Emission Trading Law and the National Carbon Emission Trading quota Allocation Valve to ensure that the carbon emission rights market obtains sufficient legal guarantees. According to the current legislative plan, the drafting and demonstration of the National Carbon Emission Trading Law is still slow, but departmental regulations lack legal effect. To only stipulate the authority of the competent department over the management of the relative person's behavior, no new rights were set up. There is a lack of punishment. The maximum penalty for the subject of the carbon emission obligations that fails to fulfill the settlement obligation is only 30,000 yuan. Therefore, based on fully absorbing the opinions of all parties on the Interim Regulations on Carbon Emissions Trading, we should introduce the interim regulations as soon as possible. Then, we should learn more useful experience on this basis, formulate a special National Carbon Emission Trading Law to protect the rights of all parties. On this basis, we should continue to formulate the National Carbon Emission Trading Quota Allocation Law, reasonably define the boundaries of power of the legislature and administrative organs in the quota allocation, and expand the scope of its main carbon emission obligations. Going forward, we will further improve the level and quality of legislation.

---

62  Fu, Shibo, et al. "Spatiotemporal Evolution and Improvement Pathways of Sustainable Entrepreneurship Efficiency: Alleviating Environmental and Employment Pressures." Available at SSRN 5110386.
63  Song X (2021) BinHui: from local carbon pilot to national carbon trading center. China Environ Manag.

*6.2.2 Formulate judicial interpretations to promote carbon-quota trading among enterprises*

Article 509 of the Civil Code Notwithstanding the principled provision that the performance of the contract shall comply with the green principle, but being excessively vague and unspecific, making it is operable in practice. The interpretation of this clause is necessarily involved in carbon financial transactions. However, since the clause does not specify how green principles can be reflected in reducing carbon emissions in financial transactions, in practice, if there are such disputes, the judge will face an embarrassing situation. However, the revision procedure of the civil code is complicated, making it difficult to fill the legal loopholes in the issue in the short term, therefore, in combination with the relevant legislative documents of carbon emission right trading, targeted judicial interpretation of carbon trading, type and detail the possible breach of contract in the process of carbon trading. Thus, carbon financial trading can obtain a predictable and stable judicial guarantee. The relevant legal assumptions involved in the legal system of carbon financial trading present the relationship shown in Fig. 2.

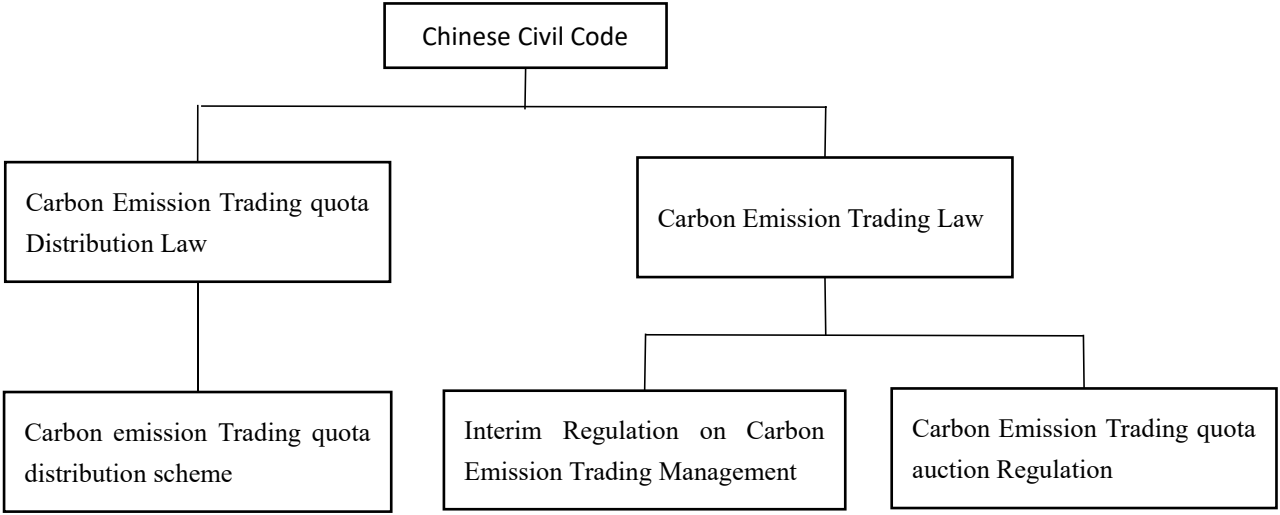

**Fig. 2** The legal system involved in carbon financial trading

*6.2.3 We will promote commercial amendments to improve environmental information disclosure mechanisms*

In the future, we should focus on building an environmental and social governance (ESG) system with Chinese characteristics and improve the environmental information disclosure mechanism. The carbon finance system, integrated with Environmental, Social, and Governance (ESG) frameworks, plays a crucial role in advancing carbon peak and carbon neutrality goals. By aligning financial investments with sustainable practices, the ESG system promotes the development of low-carbon technologies and eco-friendly industries. It encourages businesses to adopt responsible environmental practices, reducing carbon emissions and optimizing resource use. Through mechanisms such as green bonds and carbon credits, the system helps channel funds into projects that support the transition to a low-carbon economy, thereby contributing to the achievement of dual carbon goals.[64] As an important supporting venue for the two-carbon target, ESG is an important step to improve the carbon financial system. Therefore, the ESG concept should be implemented at the level of commercial law, and the commercial legal system should be improved to inject source water into the development of carbon finance. Specifically, at the level of financial regulation, on the one hand, the inclusion of ESG into the industry standard improves the ESG performance of the industry, promotes the improvement of energy efficiency in the industry, and reduces the carbon emission level and carbon neutral risk of the industry; on the other hand, through the normative role of the regulatory authorities, unified ESG industry information disclosure standards are an important means to effectively promote the process of carbon neutrality.[65] It is suggested that the future should build the ESG index system with Chinese characteristics, refer to the equator principle of the "gold criterion" concept, tools and methods, establish a setup from the decision to execution, from the system to the process of ESG index management system, effective disclosure of carbon financial information, and at the same time, according to the international ESG information disclosure standards disclosure principle, improve the law, the carbon finance law, the green climate finance law, revise the commercial bank law, the company law, the securities law, the insurance law, the trust law and other commercial law, and support the introduction of ESG concept. The perfect legal guarantee system in the field of carbon

---

64 Mei, Ke, Jiaying Zhang, and Jinshan Guo. "Mechanism of ESG System Construction Promoting Sustainable Financial Development under the" Dual Carbon" Goals." Advances in Economics, Management and Political Sciences 119 (2024): 58-64.
65 Shi Y, Deng J, Lou YT, Jin L (2021) ESG development points to promote carbon neutralization. New Financ Manag.

financial financing in the future is shown in the Fig 3.

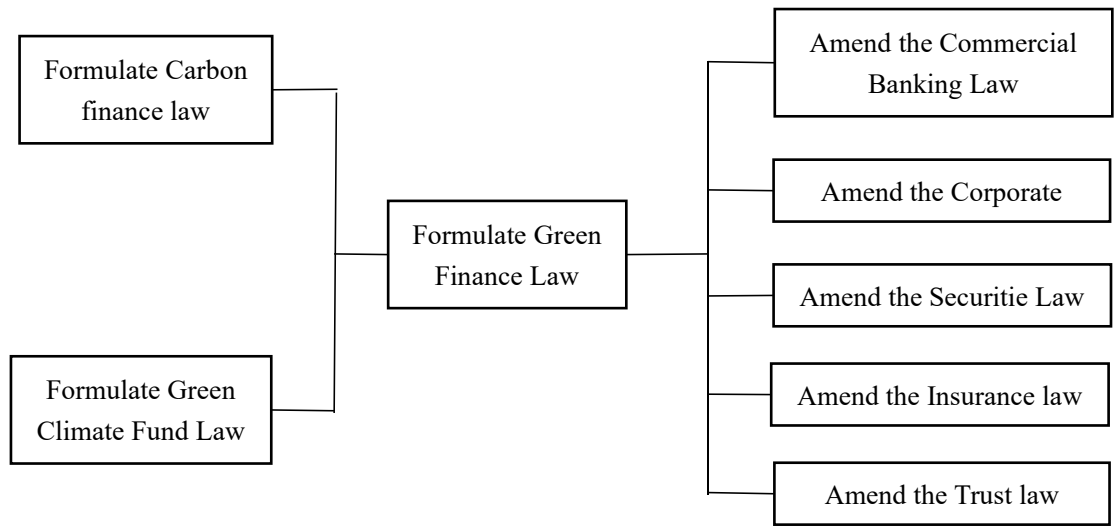

**Fig. 3** The legal guarantee system in the field of carbon finance financing

*6.2.4 In the future, we will promote the linkage mechanism between environmental judicial specialization and carbon finance case hearing*

At present, the dual-carbon target promotes the rapid development of the carbon financial market, with carbon-related disputes increasing. From the perspective of regulating the carbon financial market, it is not enough to fully protect the orderly development of carbon finance, as it is difficult to promote the realization of in-depth goals such as the transformation of the production mode. Therefore, the environmental judicial level must solve the environmental resources judicial system of the double carbon target judicial security problem. The key to solving the problem is in environmental judicial specialization, only by involving carbon dispute-specific provisions and elaboration on the specific types of cases, and foster a large number of proficiency in carbon-related business environment resources trial judges, effectively solving the involved carbon dispute cases, providing reasoning on rough problems. The jurisdiction of carbon trading and carbon finance cases should be clarified as soon as possible, carbon finance cases that seriously endanger ecological environmental protection should be submitted to the environmental and resources trial court for trial, and environmental public interest litigation cases related to carbon finance cases should be heard together. However, carbon finance cases that do not reach the above severity should be heard by green finance courts. Based on the relevant trials of green finance courts in Guizhou that have good results, it is suggested that Shanghai set up a green finance court based on the infrastructure advantages of the national carbon trading platform and Shanghai Financial Court to hear such carbon finance cases.

# 7 Epilogue

In the next 30 years, the investment scale of the carbon neutral field is expected to exceed 138 trillion yuan, and the scale of fiscal expenditure may only account for 15.94%.[66] Carbon finance is the main way to pay for the funding gap of the carbon neutral economy. Summarizing the development status of the past carbon financial system, it is found that the obstacles to the development of the carbon finance system lie in the unreasonable economic structure, the insufficient construction of the carbon finance market, the single type of carbon finance products, the weak willingness of enterprises to use carbon finance, and the pending judicial guarantee system of carbon finance. Therefore, the system should be improved specifically from the economic level and the legal level. Economic countermeasures include adjusting the structure of economic development, strengthening the infrastructure construction of the carbon finance trading market, encouraging financial institutions to develop and pilot more carbon financial products, and strengthening the publicity and education of carbon finance. Legal countermeasures include raising the legislative level of carbon quota allocation, formulating judicial interpretation to promote carbon quota trading among enterprises, promoting commercial law amendment and improving the environmental information disclosure mechanism, and promoting the linkage mechanism between environmental judicial specialization and the trial of carbon financial cases. To sum up, carbon finance is the key to double carbon goals, including financial support, and funding of the carbon peak. Carbon neutral system construction is the water of the source, which acts only by building and perfecting China's carbon financial system, innovation and development of carbon financial judicial practice, to ensure stable and orderly healthy development of the carbon financial system, forming the overall pattern of pollution reduction and carbon reduction synergies.

---

66 Wang, Chunshu, Qiong Wang, and Zhifeng Yue. "Research on the Impact of Green Finance on Industrial Structure in Shaanxi Province." Academic Journal of Business & Management 6.5 (2024).

**List of Abbreviations**

Not applicable.

**Declarations**

*Ethics approval and consent to participate*

Ethics approval and consent to participate was not required for this research.

*consent for publication*

All authors approved the final manuscript and the submission to this journal.

*Availability of data and material*

All relevant data are within the paper.

*Competing interests*

The authors declare that they have no conflict of interest.

*Funding*

This study was supported by Research on the Theory of Compensation for Ecological Environment Damage, WEH3457018.

*Authors' contributions*

ZYD designed the study, HH collected the material and analyzed the data, and was a major contributor in writing the manuscript. All authors read and approved the final manuscript.